\documentclass[prl,nofootinbib,twocolumn,superscriptaddress,preprintnumbers,balancelastpage,nofootinbib]{revtex4-1}

\usepackage{mciteplus}
\usepackage{latexsym}
\usepackage{amsthm}
\usepackage{amsmath}
\usepackage{amssymb}
\usepackage{hepunits}
\usepackage{hyperref}
\usepackage{bbm}
\usepackage{bm}
\usepackage{xfrac}
\usepackage{color}
\usepackage[dvipsnames]{xcolor}
\usepackage{colordvi}
\usepackage{comment}
\usepackage{dcolumn}
\usepackage{times,latexsym,graphicx,wrapfig,url}
\usepackage{epsfig,lineno,bm}
\usepackage{subfigure}
\usepackage{slashed}\usepackage{graphicx}
\usepackage{amssymb}
\usepackage{bm}
\usepackage{mathrsfs}
\usepackage{amsmath}
\usepackage{amsfonts}
\usepackage{color}
\usepackage{tikz}
\usepackage{slashed}
\usepackage{cancel}
\usepackage{hyperref}
\usepackage{slashed}
\usetikzlibrary{arrows}
\usetikzlibrary{shapes}
\usetikzlibrary{trees}
\usetikzlibrary{matrix}
\usetikzlibrary{arrows}

\newcommand{\be}{\begin{equation}}
\newcommand{\ee}{\end{equation}}
\newcommand{\bea}{\begin{eqnarray}}
\newcommand{\eea}{\end{eqnarray}}

\newcommand{\bef}{\begin{figure}[htbp]\begin{center}}
\newcommand{\eef}{\end{center}\end{figure}}
\def\lsim{\mathrel{\rlap{\lower4pt\hbox{\hskip1pt$\sim$}}
    \raise1pt\hbox{$<$}}}
\def\gsim{\mathrel{\rlap{\lower4pt\hbox{\hskip1pt$\sim$}}
    \raise1pt\hbox{$>$}}} 

\newcommand\Princeton{ Princeton University, Princeton, NJ }
\newcommand\FNAL{Fermi National Accelerator Laboratory, Batavia, IL }
\newcommand\UMich{University of Michigan, Ann Arbor, MI }
\newcommand\KICP{Kavli Institute for Cosmological Physics, University of Chicago, Chicago, IL USA}
\newcommand\UIUC{University of Illinois Urbana-Champaign, Urbana, IL USA}

\def\lsim{\mathrel{\rlap{\lower4pt\hbox{\hskip 0.5 pt$\sim$}}
\raise1pt\hbox{$<$}}}

\newcommand{\nue}{\nu_e}
\newcommand{\nuebar}{\overline{\nu}_e}
\newcommand{\MB}{MiniBooNE}

\begin{document}

\preprint{FERMILAB-PUB-18-295-A, PUPT 2566}
\vspace{1cm}

\title{Severe Constraints on New Physics Explanations of the MiniBooNE Excess}

\preprint{}

\author{Johnathon R. Jordan}
\email{jrlowery@umich.edu}
\affiliation{\UMich}
\author{Yonatan Kahn}
\email{ykahn@uchicago.edu}
\affiliation{\Princeton}
\affiliation{\KICP}
\affiliation{\UIUC}
\author{Gordan Krnjaic}
\email{krnjaicg@fnal.gov}
\affiliation{\FNAL}
\author{Matthew Moschella}
\email{moschella@princeton.edu}
\affiliation{\Princeton}
\author{Joshua Spitz}
\email{spitzj@umich.edu}
\affiliation{\UMich}

\begin{abstract}

The MiniBooNE experiment has recently reported an anomalous 4.5$\sigma$ excess of electron-like events consistent with $\nu_e$ appearance from a $\nu_\mu$ beam at short-baseline.  Given the lack of corresponding $\nu_\mu$ disappearance observations, required in the case of oscillations involving a sterile flavor, there is strong motivation for alternative explanations of this anomaly. We consider the possibility that the observed electron-like  
signal may actually be initiated by particles produced in the MiniBooNE beryllium target,
without involving new sources of neutrino production or any neutrino oscillations. We find that the electron-like event energy and angular distributions in the full MiniBooNE data-set, including neutrino mode, antineutrino mode, and beam dump mode, 
 severely limit, and in some cases rule out, new physics scenarios as an explanation for the observed neutrino and antineutrino mode excesses. Specifically, scenarios in which the particle produced in the target either decays (visibly or semi-visibly) or scatters elastically in the detector are strongly disfavored. Using generic kinematic arguments, we extend the existing MiniBooNE results and interpretations to exhaustively constrain previously unconsidered new physics signatures, and emphasize the power of the MiniBooNE beam dump search to further constrain models for the excess.

\end{abstract}

\maketitle

%%%%%%%%%%%%%%%%%%%%%%%%%%%%%%%%%%%%%%%%
%%%%%%%%%%%%%%%%%%%%%%%%%%%%%%%%%%%%%%%%

%                                              Section I: Introduction     				        %

%%%%%%%%%%%%%%%%%%%%%%%%%%%%%%%%%%%%%%%%
%%%%%%%%%%%%%%%%%%%%%%%%%%%%%%%%%%%%%%%%

The MiniBooNE and MiniBooNE-DM experiments detect neutrinos and antineutrinos created in the Booster Neutrino Beamline at Fermilab using a 445~ton fiducial volume mineral oil detector surrounded by 1280 photomultiplier tubes~\cite{AguilarArevalo:2008qa}. The beamline is capable of operating in three distinct modes: neutrino, antineutrino, and beam dump. In neutrino and antineutrino modes, the beams originate from the production and subsequent decay of parent mesons ($\pi^\pm$ and $K^\pm$) created in the interaction of 8~GeV protons with the beryllium target. An electromagnet focuses positive (negative) parent mesons, which usually decay-in-flight (DIF) in the downstream decay-pipe, to produce the neutrino (antineutrino) beam. In beam dump mode, the proton beam is directed off target to an absorber (dump) downstream, effectively limiting the production of DIF neutrinos and antineutrinos, which can act as backgrounds for exotic searches.

\MB\ has recently reported a $4.5\sigma$ excess ($381.2 \pm 85.2$ events) of $\nu_e$-like events in a neutrino mode search for $\nu_\mu \to \nu_e$ oscillations via charged-current quasi-elastic (CCQE; $\nu_e n \to e^- p$) scattering inside the detector~\cite{Aguilar-Arevalo:2018gpe}.
Combining this neutrino mode result with the related $2.8\sigma$ ($79.3 \pm 28.6$ events) \MB\ excess in antineutrino mode \cite{Aguilar-Arevalo:2013pmq} increases the significance of the anomaly to $4.8\sigma$. (We consider the neutrino mode results as representative of both neutrino and antineutrino mode in this Letter.) If interpreted as evidence of a sterile neutrino in a 3+1 mixing scenario, these measurements are also consistent with the longstanding  $3.8\sigma$ excess in $\nuebar$ appearance 
reported by the LSND collaboration~\cite{Aguilar:2001ty};
combining LSND and \MB \ results yields a $6.1\sigma$ discrepancy with respect to the 3$\nu$ Standard Model (SM) prediction.

However, in light of several null oscillation results (for example, see Refs.~\cite{Armbruster:2002mp,Abe:2014gda, Adamson:2011ku,Cheng:2012yy, TheIceCube:2016oqi}), especially those with sensitivity to $\nu_\mu$ disappearance, the standard 3+1 sterile neutrino hypothesis is strongly disfavored~\cite{Dentler:2018sju} and adding additional sterile flavors only marginally reduces this tension \cite{Dentler:2018sju,Collin:2016rao}. Given these constraints, there is significant interest in alternative interpretations of the MiniBooNE excess beyond models of neutrino oscillations with additional sterile states. 

Adding to this motivation, in recent years there has been considerable attention devoted to exploring light weakly-coupled particles at neutrino experiments. Indeed, such experiments are powerful probes of dark matter (DM) \cite{deniverville:2012ij,deNiverville:2016rqh,Izaguirre:2017bqb,Kahn:2014sra,Jordan:2018gcd}, dark photons \cite{Park:2017prx}, millicharged particles \cite{Magill:2018tbb}, and light scalars \cite{Izaguirre:2014cza,Pospelov:2017kep}, to name just a few popular examples (see \cite{Battaglieri:2017aum,Alexander:2016aln,Essig:2013lka} for reviews). Thus, it is natural to wonder whether the \MB~  excess could be explained with light new particles which decay or scatter inside the detector to mimic a $\nu_e$ CCQE-like event satisfying \MB~ signal criteria. 

In this Letter, we critically examine the possibility that new particles beyond the SM could explain the anomalous \MB\ excess, without invoking neutrino oscillations. 
We perform a model-independent analysis of decay and scattering signatures initiated by physics unrelated to neutrino production and propagation, and reach two main conclusions, based on kinematic arguments alone: 

\begin{enumerate}

\item The simplest new physics scenarios that can produce events appearing as $\nu_e$-like are characterized by a single reconstructed electromagnetic (i.e.\ electron or photon, which are indistinguishable at \MB) track with visible energy $E_e > 140$ MeV, involving either
	\begin{enumerate}
\item  a new unstable particle produced in the target, decaying visibly- or semi-visibly in the detector to near-overlapping daughters reconstructed as a single track, or

\item a (quasi-)stable particle produced in the target, which elastically scatters 
off a detector electron,
\end{enumerate}
can in principle accommodate the reconstructed energy distribution of the excess. However, these scenarios are broadly ruled out by the measured angular distribution of the tracks constituting the excess. This distribution contains a significant
number of electromagnetic tracks with $\cos \theta_e < 0.8$ and neither of these simple processes can explain this feature.

\item In principle, (quasi-)stable particles produced in the target which undergo {\it inelastic} processes with detector nuclei and create electromagnetic tracks (e.g.\ by upscattering to a heavier state that decays to collimated $e^+e^-$ pairs or $\gamma \gamma$, or by absorption followed by emission of a photon) can explain both the energy and angular distributions, since the kinematics are very similar to true $\nu_e$ CCQE.
 However, new particle production from neutral meson decays or continuum processes is ruled out by the beam dump search from \MB\ \cite{Aguilar-Arevalo:2018wea}, which observed only two events in their electron-like signal region ($\cos \theta_e>0.9$ and $75~\mathrm{MeV}<E_{\rm vis}<850~\mathrm{MeV}$), consistent with background expectations. These processes are accessible at nearly identical rates with nearly identical spectra in both neutrino and beam dump mode, so the only difference in event rate should arise from the variation in collected POT between the two runs. This explanation is inconsistent with the simultaneous observation of an excess in neutrino mode and the null result in beam dump mode.

\end{enumerate}

The only new physics models that satisfy all of these requirements involve new interactions which mimic the inelastic upscattering characteristic of CCQE kinematics, initiated either by a neutrino or by a new particle $X$ produced in \emph{charged} meson decays ($\pi^\pm$ or $K^\pm$; e.g. $K^+ \rightarrow \mu^+ \nu_\mu X$).
For the former case, one example of such a model is \cite{Bertuzzo:2018itn} in which the $\nu_\mu$ produced in 
neutrino mode couples to a light new vector and scatters inelastically off detector
nuclei; each scatter produces a new unstable particle which decays to 
collimated $e^+e^-$ pairs and fakes
the signal $\nu_e$ CCQE electron. For the latter case, we do not attempt to build an explicit model, but point out that DM produced only in charged meson decays at beam dumps is a novel scenario not typically considered in the literature. In either case, if the excess is due to new physics, a signal is guaranteed in \MB's beam dump mode with sufficiently high exposure due to the similarity of the charged meson spectra shapes as compared to neutrino mode, so a null result with a modest improvement in POT would suffice to rule out these remaining explanations.

This Letter is organized as follows. We first 
apply constraints from the angular distribution of the MiniBooNE excess to rule out decay and elastic scattering explanations. We then use the null
results from the \MB\ beam dump search to identify the conditions under which inelastic signatures can be consistent with the data. Finally, we summarize our findings and offer some concluding remarks.

%%%%%%%%%%%%%%%%%%%%%%%%%%%%%%%%%%%%%%%%
%%%%%%%%%%%%%%%%%%%%%%%%%%%%%%%%%%%%%%%%

%                     Section II:    Angular Distribution Constraint      		            	 %

%%%%%%%%%%%%%%%%%%%%%%%%%%%%%%%%%%%%%%%%
%%%%%%%%%%%%%%%%%%%%%%%%%%%%%%%%%%%%%%%%

\begin{figure}[t!]
\hspace{-0cm}
\includegraphics[width=0.46\textwidth]{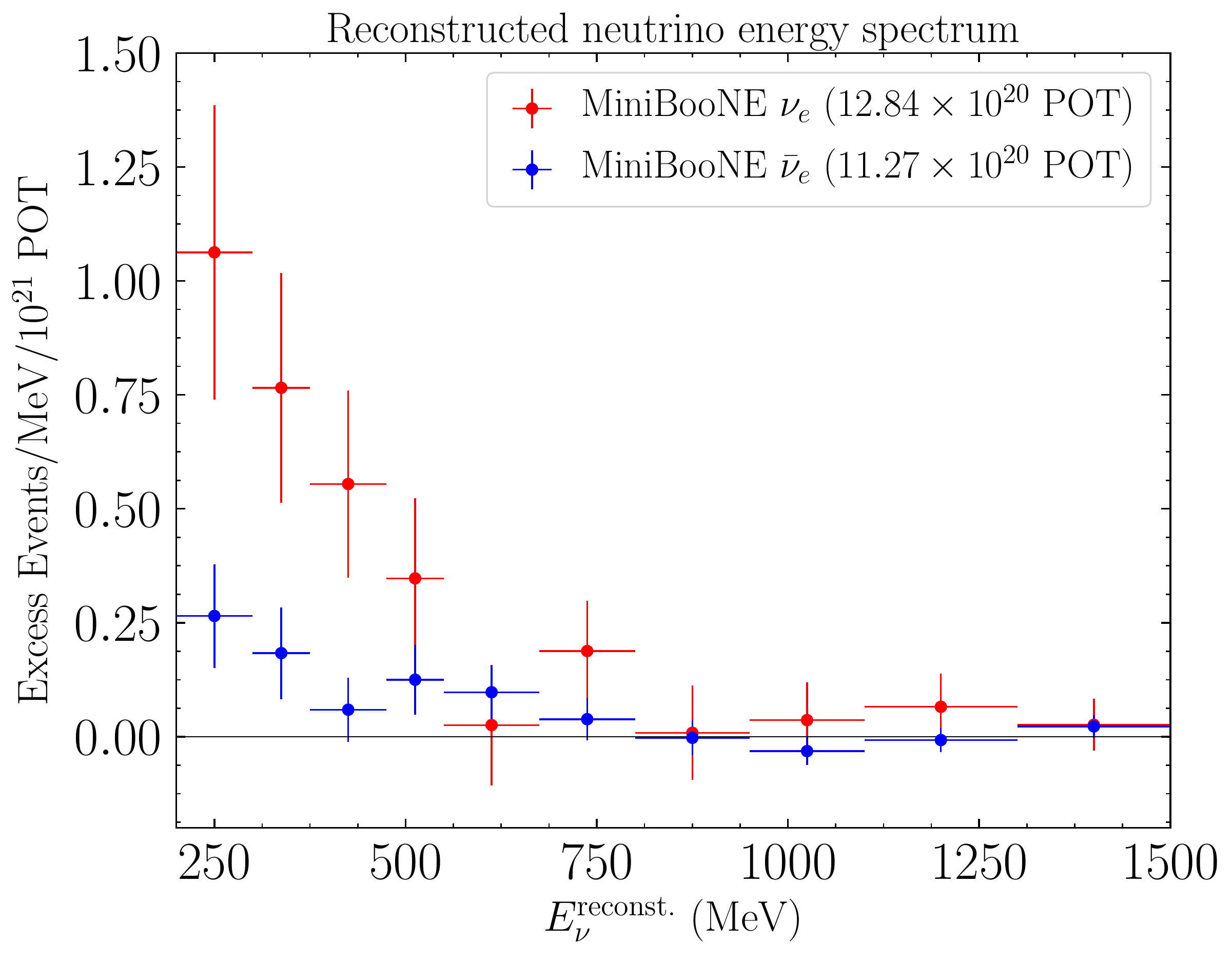}
\caption{ 
The energy spectra of the \MB \ excesses in neutrino mode (red) \cite{Aguilar-Arevalo:2018gpe} 
and antineutrino mode (blue) \cite{Aguilar-Arevalo:2013pmq}
 presented in terms of  
the reconstructed neutrino energy $E_\nu^{(\rm reconst.)}$ from 
Eq.\ (\ref{eq:Erecon}). For the simple decay and elastic scattering scenarios described in the text, we find that even when these processes 
accommodate the energy distributions here, they fail to match the angular
profile of the excess in Fig.~\ref{FIG:CosTheta}. For the inelastic
signatures described in the text, it is possible, in principle, to accommodate both  the energy distribution here and the angular distribution in Fig.~\ref{FIG:CosTheta}, but 
such scenarios are strongly constrained by the null results from the \MB \ beam
dump search in \cite{Aguilar-Arevalo:2018wea}.
} 
\label{FIG:Spectrum}
\end{figure}

\begin{figure}
\hspace{-0cm}
\includegraphics[width=0.44\textwidth]{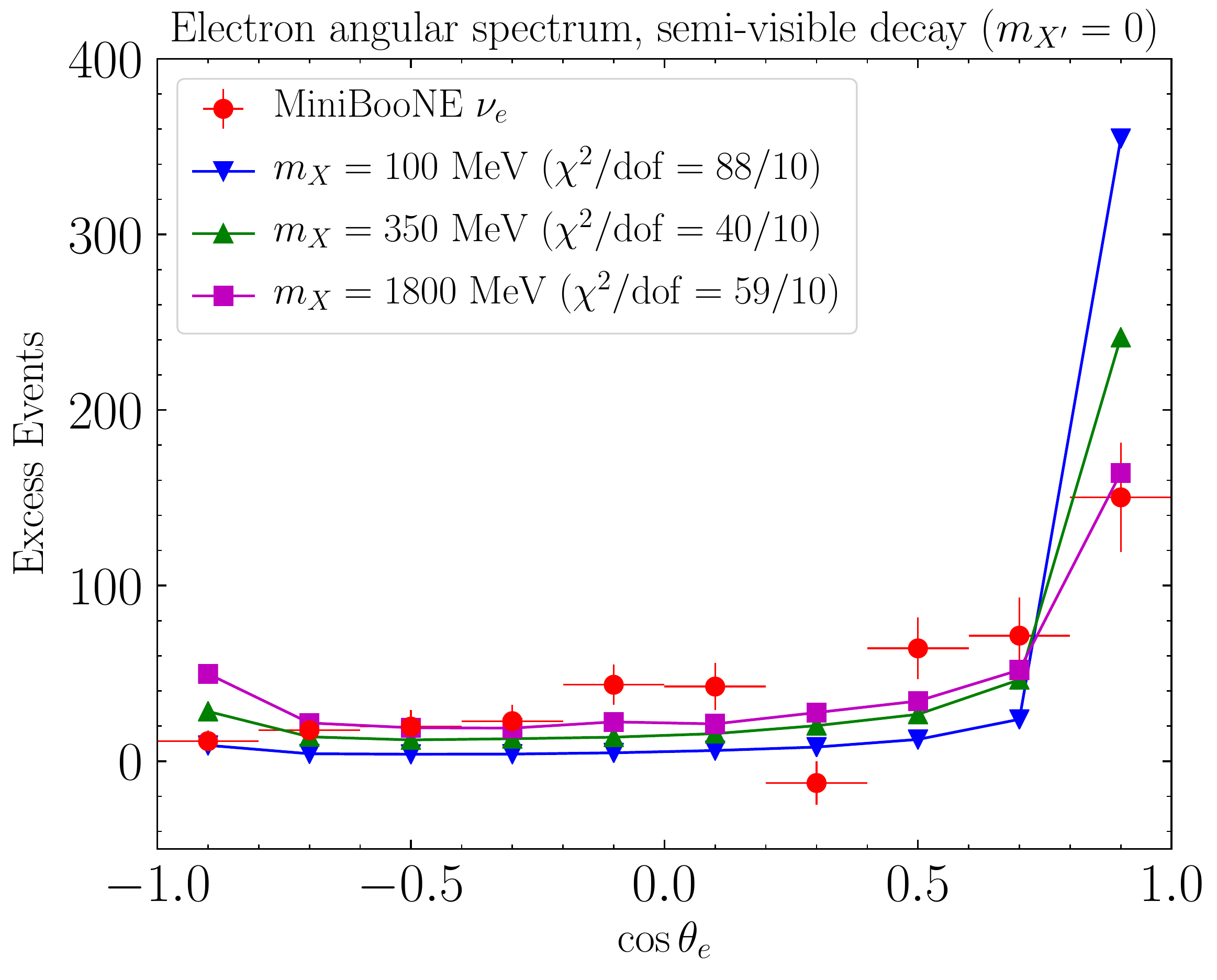}
\caption{Angular distributions for semi-visibly decaying particles $X \to X' + p_{\rm EM}$ , where $X^\prime$ is invisible (with $m_{X'} = 0$) and $p_{\rm EM}$ is 
the electromagnetic energy deposition in the MiniBooNE detector, which is chosen to fit the energy distribution
of the neutrino mode excess shown in Fig.~\ref{FIG:Spectrum}. The energy spectrum of $X$ determines its boost and angular
distributions, the latter of which is plotted here for various $m_X$. The observed angular distribution of the $\nu_e$ excess from \cite{Aguilar-Arevalo:2018gpe} is shown in red. Note that choosing
$m_{X^\prime} = 0$ here is conservative because heavier $X^\prime$ make it more difficult to generate wide-angle visible energy. The semi-visible decay hypothesis is strongly disfavored by the angular distribution regardless of $m_X$, as indicated by the $\chi^2$ values (based on statistical errors only) for each mass.
}
\label{FIG:CosTheta}
\end{figure}

{\it Angular Distribution Constraints.} We consider three candidate detector signatures from new particles produced in the \MB\ target: unstable 
particles that decay to fully visible final states, unstable particles decaying to semi-visible final states, and stable particles
that scatter elastically off detector electrons. The event selection criteria for the $\nue$ CCQE search seeks to isolate single electron-like tracks with reconstructed neutrino energy 200 MeV $< E_\nu^{(\rm reconst.)} < 1250$ MeV, determined from the outgoing electromagnetic track energy $E_e$ and angle $\theta_e$ with respect to the beam axis by
\be
\label{eq:Erecon}
E_\nu^{(\rm reconst.)} = \frac{2m_n E_e + m_p^2 - m_n^2 - m_e^2}{2(m_n - E_e + \cos \theta_e \sqrt{E_e^2 - m_e^2})}.
\ee
The measured spectra of $E_\nu^{(\rm reconst.)}$ for the neutrino and antineutrino mode excesses is shown in Fig.~\ref{FIG:Spectrum}, and the angular spectrum $\cos \theta_e$ in neutrino mode is shown in red in Fig.~\ref{FIG:CosTheta}. 

For both new particle visible decay and elastic scattering, we will show that even when the electromagnetic energy  
 deposited inside the detector mimics the $\nu_e$ CCQE signal and accommodates the energy spectrum of the excess,  the corresponding
angular spectrum is always forward-peaked ($\cos\theta_e > 0.99$), and therefore unable to explain the observed angular distribution. The semi-visible decay scenario is able to produce wider-angle events but is still strongly disfavored by the shape of the angular distribution. As we will discuss further below, the null results from the beam dump search can be used to derive additional constraints on these simple models, assuming the production of new particles scales with the number of protons on target (POT), but these constraints will be somewhat model-dependent. The arguments presented in this section are insensitive to the production mechanism and based solely on kinematics.

Consider first the possibility that a long-lived neutral state $X$ is produced at the \MB\ target and survives to the detector 541~m downstream, where it decays visibly to $e^+e^-$ or $\gamma \gamma$. As has been pointed out in \cite{Bertuzzo:2018itn,Ballett:2018ynz}, overlapping electromagnetic tracks from $X$ decay can contribute to the observed excess. The reconstructed energy of the track is equal to the sum of the two track energies, $E_e = E_{1} + E_{2}$, and the precise opening angle $\theta_{12}$ at which two tracks can be distinguished depends on the details of the track reconstruction procedure \cite{Raaf:2005up,Patterson:2009ki}; in this analysis we consider two tracks indistinguishable from a single track if
\be
\cos \theta_{12} > 1 - \frac{(30 \ \MeV)^2}{2E_1 E_2},
\ee
which is motivated by MiniBooNE's use of the two-track invariant mass variable
\be
m_{\rm track} \equiv \sqrt{2E_{1}E_{2}(1 - \cos \theta_{12})},
\ee
where $\theta_{12}$ is the angle between the two tracks and $E_{1,2}$ are the track energies; if $m_{\rm track} < 30 \ \MeV$, \MB\ is unable to distinguish $e^+e^-$ or $\gamma \gamma$ pairs from single electrons \cite{Karagiorgi:2010zz}. This cut also ensures that such events do not appear in the neutral-current $\pi^0$ event sample \cite{AguilarArevalo:2009ww}. For the present scenario, visible $X$ decays would contribute to the CCQE excess if $m_X < 30 \ \MeV$. As described above, the reconstructed track angle is weighted by the track energies; by momentum conservation, this sum is simply the original $X$ 4-vector, which must satisfy $\cos \theta_e > 0.9999 $ in order for $X$ to enter the \MB\ detector, a sphere of fiducial radius 5.75 m located 541 m away from the target. This is highly inconsistent with the $\cos \theta_e$ distribution of the excess (see Fig.~\ref{FIG:CosTheta}), which shows significant contributions from $\cos \theta_e < 0.8$. In particular, a model which matches the size of the neutrino mode excess (381.2 events), but predicts all events to have $\cos \theta_e > 0.8$ is incompatible with the observed excess of 150 $\pm$ 31 in this bin (in consideration of statistical errors only; systematics and bin-to-bin correlations are not available, noting that the angular resolution is 3-5$^\circ$ for 100-600~MeV electron energies in $\nu_e$ CCQE events~\cite{Patterson:2009ki}).

Since new particles with fully visible decays necessarily give forward-peaked energy depositions in
conflict with the angular distribution of the measured excess, we now consider the possibility
that the decay signature of a new unstable particle $X$ features partially invisible final states. Decays to a lighter  invisible state $X'$ plus a single charged particle are forbidden by charge conservation, and $X \to X' + \pi^0$, where the resulting $\gamma \gamma$ is reconstructed as a single track, would create a large two-track excess not seen in the neutral-current $\pi^0$ analysis~\cite{AguilarArevalo:2009ww}. Thus, the dominant allowed channel is a two-body decay where $X$ decays into a lighter dark-sector state $X'$ and a photon ($X \to X' + \gamma$). Since the electromagnetic tracks must be well-collimated to contribute to the excess, our argument applies equally well to decays with more final state particles than just a single photon, where $\gamma$ now represents collimated electromagnetic energy. In what follows we will treat this scenario as a quasi-two-body decay, where the electromagnetic energy is considered as a single 4-vector $p_{\rm EM}$ with $0 \leq p_{\rm EM}^2 \leq (30 \ \MeV)^2$.  

In the $X$ rest frame, the electromagnetic energy is $E_{\rm EM} = (m_X^2 - m_{X'}^2)/2m_X$. Electromagnetic energy with small invariant mass compared to the beam energy, emitted backwards in the $X$ rest frame, will be boosted to very small lab-frame energies,
\be
\label{eq:Elab}
E_{\rm EM, \ lab} \approx \frac{m_X^2 - m_{X'}^2}{2m_X} \gamma(1-\beta),
\ee
where $\gamma$ and $\beta$ are the boost and velocity of $X$, respectively. This will make it difficult for such an event to pass the $E_e > 140$ MeV selection for the $\nu_e$-like excess unless the mass splitting between dark states $m_X^2 - m_{X'}^2$ is large to make up for the (typically very small) $1 - \beta$ factor.

To illustrate the difficulty, we perform a Monte Carlo simulation of $X \to X' + p_{\rm EM}$ where the boost distribution of $X$ is constructed to \emph{exactly} reproduce the observed $\nu_e$ excess shown in Fig.~\ref{FIG:Spectrum}. Since unpolarized two-body decays are only a function of the invariant masses $m_{X'}^2$ and $p_{\rm EM}^2$, the matrix element is isotropic in the rest frame,\footnote{If parity is violated in $X$ decays, the matrix element may be anisotropic. However, this effect does not change any of our qualitative
conclusions requiring $m_X \gg m_{\pi^0}, m_K$ in order to improve the fit to the angular distribution. Since such heavy particles
 can only arise in continuum processes at MiniBooNE (e.g. proton bremsstrahlung), this possibility is strongly constrained by
the beam dump results from Ref. \cite{Aguilar-Arevalo:2018wea} discussed below. } so sampling from a uniform angular distribution in the $X$ rest frame and boosting according to the distribution inherited from the energy spectrum determines the lab-frame kinematics. We see that for $m_{X'} = 0$, the angular distributions are discrepant with large $\chi^2$, regardless of $m_X$. The minimum $\chi^2$ value as a function of $m_X$ was found for $m_X = 350 \ \MeV$, which is still strongly disfavored with $\chi^2/{\rm dof} = 40/10$. Attempting to match the highest $\cos \theta_e$ bin only increases the discrepancy in the lowest $\cos \theta_e$ bin, as shown by the $m_X = 1800 \ \MeV$ points.

The challenge of reproducing the angular distribution is only exacerbated for $m_{X'} > 0$ by the arguments surrounding Eq.~(\ref{eq:Elab}). These kinematic arguments hold even if multiple invisible daughters $X'_i$ are produced in $X$ decay, since the invariant mass of the final-state invisible 4-vector must be non-negative. We thus conclude that semi-visible decays are also inconsistent with the \MB\ data.

Finally, we consider the possibility that the excess could be due 
to new particles that are stable on beamline length scales. Such 
particles could be produced in proton-target interactions and subsequently scatter in the detector, mimicking the $\nu_e$ CCQE signal.
Here we find that if these signatures arise from elastic scattering, there 
is generic tension with the angular distribution of the excess. 

For a (quasi-)stable new particle $X$ of mass $m_X$ to interact elastically and mimic a $\nu_e$ 
CCQE signature, it must
scatter off electrons via $Xe^- \to Xe^-$ to yield visible electromagnetic energy
in the final state. For an incident $X$ with energy $E_X$, the track angle $\theta_e$ of the 
scattered electron with total track energy $E_e$ is uniquely specified by the masses
and energies
\be \label{eq:cos}
\cos \theta_{e} = \frac{  E_{X} E_{e}  - m_e ( E_{X} + m_e -  E_{e})    }{\sqrt{( E_{X}^2  -m_X^2  )(    E^2_{e} -m_e^2 )}} ,
\ee
so in the relativistic limit, which is appropriate for beam dump mode where
  $m_X \ll E_X$ and $m_e \ll  E_e$, we have 
  \be\label{eq:cos-approx}
  \cos\theta_e =1 - m_e\left( \frac{E_X - E_e}{E_X E_e}   \right) + {\cal O} \!\left( \frac{m_e^2}{E_e^2}\right).
  \ee
  Since MiniBooNE's selection requires track energies $E_e > 140 \ \MeV$ and $E_X > E_e$ by energy conservation, this 
  process always yields $\cos\theta_e > 0.99$ with no support for tracks with $\cos\theta_e < 0.8$, which are required to explain the excess. Thus, obtaining a broader angular distribution requires scattering off nucleons or nuclei. However, elastic $X$ scattering off a nucleus does not yield any final state electrons, so any viable scattering-based 
explanation must generate the signal through inelastic electron or photon production to mimic a $\nu_e$ CCQE final state.

In principle the above argument could fail if $X$ is non-relativistic and the expansion in Eq.~(\ref{eq:cos-approx}) is not valid.
However, in order for elastic $Xe^-$ scattering to produce a wide angle electron track $\cos \theta_e \sim 0$, the numerator of Eq.~(\ref{eq:cos}) requires $E_e \sim m_e$, which is inconsistent
with the $E_e > 140 \ \MeV$ selection requirement in neutrino mode. 
 Note that even if the selection criteria were more permissive, for typical \MB \ beam energies $X$ can only be non-relativistic for $m_X \gtrsim \GeV$, and 
 therefore must be produced by a continuum process (rather than from meson decay), which is ruled out by the beam dump constraints discussed below.

%%%%%%%%%%%%%%%%%%%%%%%%%%%%%%%%%%%%%%%%
%%%%%%%%%%%%%%%%%%%%%%%%%%%%%%%%%%%%%%%%

%                     Section III:    Beam Dump Constraints      		            	 %

%%%%%%%%%%%%%%%%%%%%%%%%%%%%%%%%%%%%%%%%
%%%%%%%%%%%%%%%%%%%%%%%%%%%%%%%%%%%%%%%%

{\it Beam Dump Constraints.} We now consider a variation on the scattering signature discussed above. 
As noted in \cite{Bertuzzo:2018itn}, for example, (quasi-)stable particles undergoing
{\it inelastic} processes involving nuclei can explain the observed angular distribution. However,
for this scenario, there are significant constraints from the \MB\ 
beam dump search, the null result of which rules out this explanation for the excess unless the production of these new particles can be somehow greatly suppressed in beam dump mode compared to neutrino mode.

The \MB-DM collaboration has recently announced results from a new beam dump search for 
sub-GeV DM particles with $1.86 \times 10^{20}$ protons on target (POT) \cite{Aguilar-Arevalo:2018wea}, about a factor of 7 less than the neutrino mode run with identical proton energy. In beam dump mode, the protons are steered away from the beryllium target and delivered directly to the steel beam dump 50~m downstream. The absence of a downstream decay volume for the parent mesons in this configuration significantly reduces and softens the neutrino flux, thereby improving \MB's sensitivity to exotic, non neutrino-oscillations signatures.

The results reported in \cite{Aguilar-Arevalo:2018wea} are presented in terms of sub-GeV DM particles, which are produced in the steel beam dump through cascades from rare $\pi^0/\eta$ decays
 and proton bremsstrahlung. These processes first yield an on-shell,
 kinetically mixed dark photon $V$ (either via $\pi^0/\eta \to V \gamma$ or $pN \to pN V$ where $N$ is a target nucleus) which then decays to boosted pairs of DM particles. Finally, a DM particle enters the detector and scatters elastically with an electron to produce a final-state electron track. After applying the selection criteria outlined in \cite{Aguilar-Arevalo:2018wea}, the collaboration reports no excess signal events above background expectations. Since the selection is sufficiently similar to the neutrino mode CCQE signal definition (i.e.
one isolated electromagnetic track with $E_e > 75 \ \MeV$ and $\cos \theta_e > 0.9$), this search can be used
to place limits on any non-neutrino physics originating from the target. In both modes,
the neutral meson ($\pi^0/\eta$) distributions are similar, with negligible differences in shape arising from the composition of the target nuclei. Similarly, the bremsstrahlung distributions are expected to be identical, despite the different target compositions \cite{Blumlein:2013cua}, which informs potential production mechanisms from either ordinary or ``dark'' bremsstrahlung of a new gauge boson like $V$.

\begin{figure}
\hspace{-0.6cm}
\includegraphics[width=0.52\textwidth]{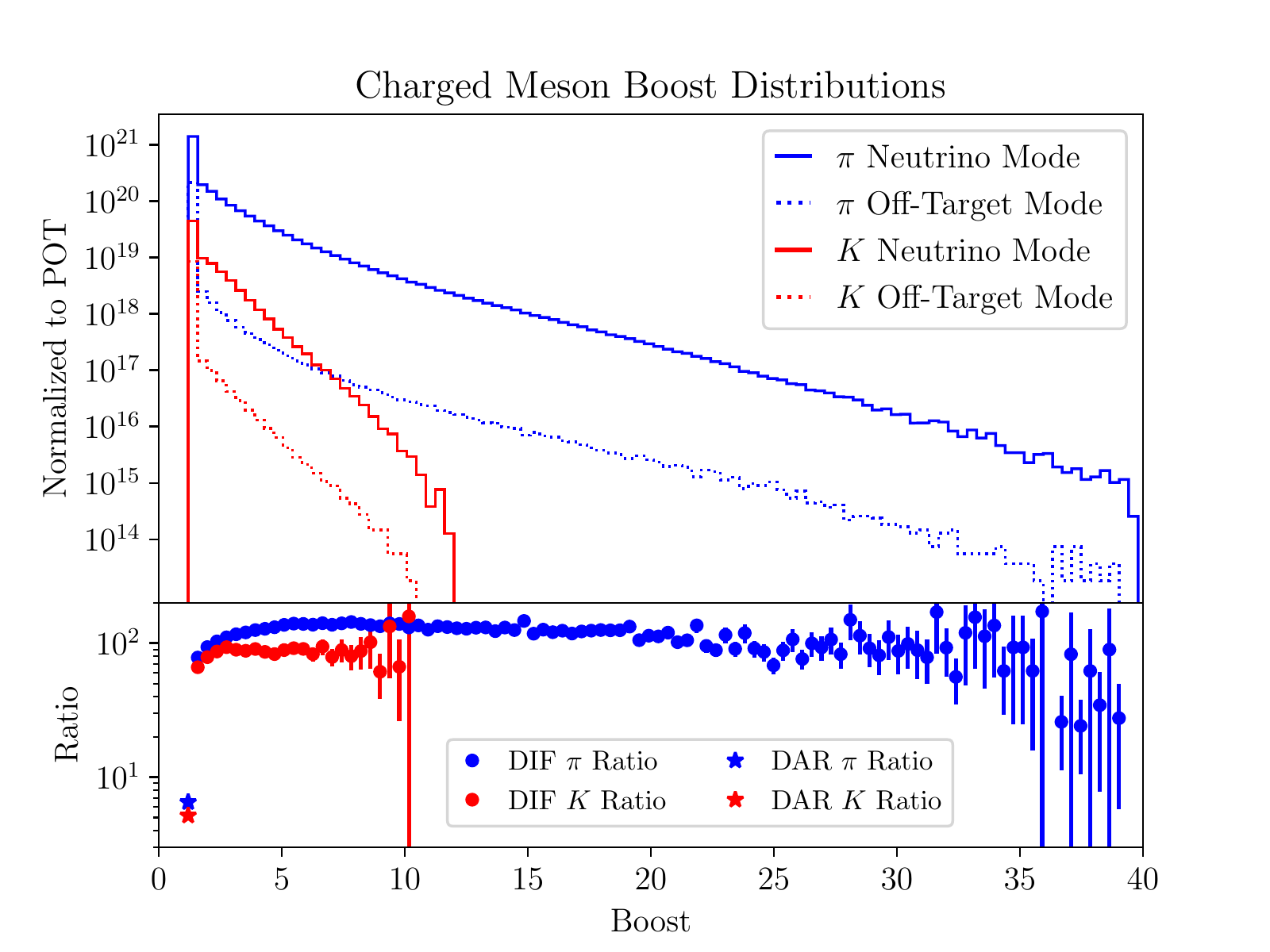}
\caption{Top: Boost distributions of charged pions (blue) and charged kaons (red) at \MB \ in both neutrino (solid) and beam dump mode (dashed). The distributions are normalized to the corresponding exposures, $12.84 \times 10^{20}$ POT for neutrino mode and $1.86 \times 10^{20}$ POT for beam dump mode. Bottom: Ratio of neutrino mode production to beam dump mode production for DIF (circles) and DAR (stars). }
\label{FIG:ChargedMesons}
\end{figure} 

This result immediately implies stringent constraints on any new particle $X$ produced from neutral meson decay or bremsstrahlung. Since the flux is essentially identical in beam dump and neutrino modes, excepting the different target-to-detector (541~m) and dump-to-detector (487~m) distances, a signal in one mode is also expected in the other, with the rate simply scaling with POT. As described in \cite{Aguilar-Arevalo:2018wea}, applying the $\nu_e$ CCQE selection to the beam dump data yields a \emph{deficit} of $2.8$ events (with 6 observed events), while an excess of $35.5 \pm 7.4$ events would have been expected based on the combined neutrino and antineutrino mode excesses. With production via neutral meson decay or bremsstrahlung therefore ruled out by $4.8\sigma$ \cite{Aguilar-Arevalo:2018wea}, if a new particle $X$ is responsible for the MiniBooNE excess, it must be created via \emph{charged} meson decay. Indeed, the size difference of the neutrino and antineutrino mode excesses might suggest a mechanism originating from charged mesons, since neutral meson and bremsstrahlung production are identical regardless of the beam magnet polarity.

Fig.~\ref{FIG:ChargedMesons} shows the Lorentz boost ($\gamma$) distributions for $\pi^{\pm}$ and $K^{\pm}$ in both neutrino and beam dump mode, normalized to the accumulated POT for each. These boost distributions were made using a detailed \texttt{Geant4}~\cite{Agostinelli:2002hh} simulation of the Booster Neutrino Beamline incorporating all relevant elements~\cite{AguilarArevalo:2008qa,AguilarArevalo:2008yp}. We see that the shapes of the DIF portions of the spectra are extremely similar, but the flux of charged mesons is reduced by a factor of $\sim$100 in beam dump mode due to the lower intergrated POT and suppressed off-target production. The boost distributions shown in Fig.~\ref{FIG:ChargedMesons} were made with the \texttt{FTFP\_BERT} physics list in \texttt{Geant4}, but we verified the similarity between the DIF spectra in neutrino mode and beam dump mode using predictions from the \texttt{QGSP\_BERT} and \texttt{QGSP\_BIC} physics lists as well.

If the kinematics of a proposed signal model require a significant boost from the charged mesons, as in the model of \cite{Bertuzzo:2018itn} where a $\nu_\mu$ upscatters to a heavy state $N_D$ with mass $\sim$320~MeV, the signal will arise from the DIF component and, with similar scaling of the background, the expected total excess in beam dump mode would be $381/100 \sim \mathcal{O}(4)$ events, with about half  of these events passing the $\cos \theta_e > 0.9$ requirement as shown in Fig.~\ref{FIG:CosTheta}. However, due to the different selection between the dark matter search in beam dump mode and the $\nu_e$ CCQE search in neutrino mode, the precise expectation for the event rate may differ by $\mathcal{O}(1)$ factors, so with the current data set it is not yet possible to exclude such explanations; these remarks apply equally well to the similar model in \cite{Ballett:2018ynz}.

On the other hand, if the signal model had lower kinematic thresholds and could explain the excess with decay-at-rest (DAR) charged mesons, for which the POT-normalized flux in the two modes only differs by a factor of $\sim$5, the signal rate would be a factor of $\sim$20 higher, and would be firmly ruled out by the beam dump search. These considerations can of course be made more precise for any particular model, but regardless of this selection, all such explanations could likely be confirmed or ruled out with an improvement of only a factor of a few in POT in beam dump mode. We emphasize that the constraints described in this section hold \emph{regardless} of whether the chosen model matches the angular distribution or not, and serve as additional constraints on the models discussed in this Letter.

{\it Conclusions.} In this Letter, we have shown that explanations for the longstanding \MB\ excess which invoke new physics are severely constrained by the angular distribution of electromagnetic tracks and the null result from the recent beam dump search. We have presented model-independent arguments against a variety of possible alternative, non neutrino-oscillations explanations, including new particle visible decays, semi-visible decays, and elastic scattering. While models invoking inelastic processes between a new (semi-)visibly decaying particle and a nuclear target, which mimic a $\nu_e$ CCQE signal, can survive the beam dump constraints, a modest increase in beam dump mode POT can rule out all such models if no excess events are observed.  We encourage the \MB\ collaboration to pursue this search with additional beam dump data~\cite{BeamDumpEOI}, which will shed further light on the source of the excess.

%Although our analysis has emphasized new particles produced in the target, we note that our conclusions also hold even if the hypothetical signal is initiated by beam neutrinos with additional (beyond SM) interactions as long as no oscillations are involved -- for examples, see the models in \cite{Bertuzzo:2018itn,Ballett:2018ynz}, which satisfy all the viability requirements outlined here. Finally, we note that, despite the historical connection between the LSND and MiniBooNE anomalies, it is generically difficult to explain the former without invoking neutrino oscillations. Unlike MiniBooNE, which triggers only on the electromagnetic energy in the final state, LSND also requires a coincident 2.2 MeV signal from neutron capture on hydrogen following the prompt positron from the inverse beta decay ($\bar{\nu}_e p \to e^+ n$), which is more difficult to mimic with other kinds of processes, including those considered here and in the specific models of  \cite{Bertuzzo:2018itn,Ballett:2018ynz}.
Although our analysis has emphasized new particles produced in the target, our conclusions also hold even in non-oscillation scenarios where the hypothetical signal is initiated by beam neutrinos with additional (beyond SM) interactions -- for examples, see the models in \cite{Bertuzzo:2018itn,Ballett:2018ynz}, which satisfy all the viability requirements outlined here. Finally, we note that it is generically difficult to explain the LSND and MiniBooNE anomalies as mutually consistent, perhaps sourced from the same new physics, without invoking neutrino oscillations. Along with the kinematic differences between the representative electron-appearance-like events ($<$53~MeV in LSND and 200-475~MeV in MiniBooNE), the signals are completely different. Unlike MiniBooNE, LSND's signal ($\bar{\nu}_e p \to e^+ n$) selection requires a double-coincidence event featuring a prompt positron followed by a 2.2~MeV neutron capture on hydrogen, which is difficult to mimic by other kinds of new physics processes, including those considered here and in the specific models of  \cite{Bertuzzo:2018itn,Ballett:2018ynz}.
 \bigskip
 
{\it Acknowledgments. ---} We thank Asher Berlin, Patrick Huber, Jia Liu, Pedro Machado, Camillo Mariani, Nadav Outmezguine, Diego Redigolo, and Tomer Volansky for helpful conversations. We especially thank Bill Louis and Richard Van de Water for enlightening discussions regarding the performance of the \MB\ detector. Fermilab is operated by Fermi Research Alliance, LLC, under Contract No. DE-AC02-07CH11359 with the US Department of Energy. JRJ is supported by the National Science Foundation Graduate Research Fellowship under Grant No. DGE-1256260. The work of YK was supported in part by the Kavli Institute for Cosmological Physics at the University of Chicago through an endowment from the Kavli Foundation and its founder Fred Kavli. JS is supported by the U.S. Department of Energy, Office of Science, under
grant DE-SC0007859.

\bibliography{MBExcessIDMbib.bib}

\end{document}